\documentclass[12pt]{article}
\usepackage{amsfonts}
\def\CellGroup{\bgroup}
\def\endCellGroup{\egroup}
\begin{document}
\begin{center}
{\large{\bf Baxter $Q$-operators for integrable DST chain.}\\}

\vspace*{10mm}

{\bf A.E. Kovalsky}\\
{\it MIPT, Dolgoprudny, Moscow reg., \\
 IHEP, Protvino, Moscow reg., Russia,} \\

{\bf G.P. Pronko}\\
{\it IHEP, Protvino, Moscow reg., Russia,\\
International Solvay Institute, Brussels, Belgium}

\begin{abstract}
Following the procedure, described in the paper \cite{TodaPronko}, for the
integrable DST chain we construct Baxter $Q$-operators \cite{Baxter} as
the
traces of monodromy of some $M$-operators, that act in quantum and
auxiliary
spaces. Within this procedure we obtain two basic $M$-operators and derive
some
functional relations between them such as  intertwining relations and
wronskian-type relations between two basic $Q$-operators.
\end{abstract} 
\end{center}

\section{Introduction}

Integrable periodic quantum DST (Discrete Self-Trapping) chain is a
quantum
system described by the following hamiltonian (it corresponds to a certain
set
of parameters $\omega_0, \gamma, \epsilon$ and $m_{ij}$ in the hamiltonian
considered in \cite{ScEil})
\begin{equation}
H=\sum_{k=1}^{N} \bigl[\varphi^+_k
\varphi_k+(\varphi^+_k\varphi_k)^2/2+\varphi^+_{k+1}\varphi_k \bigr],
\label{Hamiltonian}
\end{equation}
where the canonical variables $\varphi_k^+$ and $\varphi_k$ satisfy
commutation
relations
$[\varphi_i, \varphi_j^+]=\delta_{ij}$ and periodic boundary condintions
$\varphi_{k+N}=\varphi_k,~\varphi_{k+N}^+=\varphi_k^+$.
The system can be considered within the framework of the quantum inverse
scattering method or  $R$-matrix method. There exists the Lax operator
connected with DST chain. It acts in the tensor product of $n$-th quantum
space
and two-dimensional auxilliary space $\mathbb C^2$ (see \cite{QDST},
\cite{SklyaninDST}):
\begin{eqnarray} 
L_{n}(x)=\left(\begin{array}{cc} x-i/2-i \varphi_{n}^+ \varphi_{n} &
\varphi_{n}^{+}\\ \varphi_{n}& i  \end{array}\right),
\label{Lax}
\end{eqnarray}
where $x$ is the spectral parameter. The fundumental relations for the
matrix
elements of the Lax operator could be written in the following $R$-matrix
form:
\begin{equation}
R_{12}(x-y)L^1_n(x)L^2_n(y)=L^2_n(y) L^1_n(x) R_{12}(x-y),
\end{equation}
where the indicies 1, 2 indicate different auxiliary spaces and $R-$matrix
is
given by
\begin{equation}
R_{12}(x)=x+iP_{12},
\label{Rmatr}
\end{equation}
where $P_{12}$ is the permutation operator in the auxiliary spaces 1 and
2. The
same intertwining relations are true for the monodromy matrix
$T(x)=\prod\limits_{n=1}^N L_n(x)$ (here the multipliers are ordered from
right
to left), this in turn means that $[t(x),t(y)]=0$, where we denote
$t(x)=Tr
T(x)$ (the trace is taken over the auxiliary space). Thus  coefficients of
the
polynomials (over the spectral parameter) $t(x)$ form the family of
commuting
operators and the hamiltonian $H$ and the number of particle operator
belong to
this family (namely, if we expand $t(x)=\sum\limits_{k=0}^{N} (x-i/2)^k
H^{(k)}$, the number of particle operator $\hat n=\sum\limits_{k=1}^{N}
\varphi_k^+ \varphi_k=i H^{(N-1)}$ and hamiltonian
$H=iH^{(N-1)}-(H^{(N-1)})^2/2+H^{(N-2)}$).

The eigenvectors and the eigenvalues of $t(x)$ could be constructed in the
framework of ABA \cite{Faddeev}. 
In this approach one considers the monodromy matrix given by
$$ 
T(x)=\prod\limits_{n=1}^N L_n(x)= \left(\begin{array}{cc} \hat A(x)& 
\hat B(x) \\ \hat C(x) & \hat D(x) \end{array} \right).
$$
There exists the so-called Bethe vacuum: $C(x)\Omega=0$($\Omega=\prod\
\otimes
\omega_k$, where $\varphi_k \omega_k=0$). Vectors $\hat B(x_1)...\hat
B(x_l)\Omega$ will be eigenvectors of $t(x)=Tr T(x)=\hat A(x)+\hat D(x)$
with
eigenvalues 
$$
t(x)=(x-i/2)^N \prod\limits_{j=1}^l \frac{(x-x_j+i)}{(x-x_j)}+i^N
\prod\limits_{j=1}^l \frac{(x-x_j-i)}{(x-x_j)}
$$
provided that $x_i$ obey the Bethe equations
$$
\prod\limits_{j=1}^l
\frac{(x_i-x_j-i)}{(x_i-x_j+i)}=-\frac{(x_i-i/2)^N}{i^N}
$$
So the polynomial $q(x)=\prod_{j=1}^l(x-x_j)$ satisfies Baxter
equation:
\begin{equation}
t(x)q(x)=(x-i/2)^N q(x-i)+i^N q(x+i).
\label{BaxterEq}
\end{equation}
 According to Baxter \cite{Baxter} let us define the operator $Q(x)$ such
 that:
\begin{equation}
t(x)Q(x)=(x-i/2)^N Q(x-i)+i^N Q(x+i),
\label{Baxter}
\end{equation}
 and $[t(x),Q(y)]=0,~[Q(x),Q(y)]=0$.
 
The model under consideration occupies an intermediate place between two
other
integrable models: the XXX
spin chain and the Toda chain (Lax operators of these modeles are
intertwined
by the rational $R$-matrix (\ref{Rmatr}) too). As in the case of XXX spin
chain 
there is the $Q$-operator with polynomial eigenvalues in the spectral
parameter. It corresponds to the ABA. If we consider the equation
(\ref{BaxterEq}) as discrete analog of a second order differential
equation and
immediately there arises the question about the second solution of
(\ref{BaxterEq}). These second solutions have been intensively discussed
in
\cite{BLZ,PrStr}. The eigenvalues of the second $Q$-operator for DST model
are
meromorphic functions (in the case of Toda chain there is no ABA but there
exist two $Q$-operators: one with entire eigenvalues, second with
meromorphic
eigenvalues \cite{TodaPronko}). In the next part these two solutions of
(\ref{Baxter}) will be constructed.

The existence of the second $Q$-operator, which is linear independent from
the
first one could be seen from the following simple consideration (similar
disscussions for the case of XXX-spin chain see in \cite{PrStr}). Let us
consider Baxter equation for the eigenvalue of the first $Q$-operator -
$q(x)$,
which is a polinomial of degree $n$ (in the case of DST chain $n$ equals
the
eigenvalue of number of particle operator $\sum\limits_{i=1}^{N}
\varphi^{+}_{i}\varphi_{i}$), and the eigenvalue of the trace of monodromy
matrix $t(x)$, which is a polinomial of degree $N$:
\begin{equation}
t(x)q(x)=(x-i/2)^N q(x-i)+i^N q(x+i)
\end{equation}
or 
\begin{equation}
\frac{t(x)}{q(x+i)
q(x-i)}=\frac{(x-i/2)^N}{q(x)q(x+i)}+\frac{i^N}{q(x)q(x-i)}
\label{B1}
\end{equation}
Multiplying this equation by $\Gamma^N(-i(x-i/2))$ we get:
\begin{equation}
\frac{t(x)\Gamma^N(-i(x-i/2))}{q(x+i)
q(x-i)}=\frac{i^N\Gamma^N(-i(x+i/2))}{q(x)q(x+i)}+
\frac{i^N\Gamma^N(-i(x-i/2))}{q(x)q(x-i)}
\label{B2}
\end{equation}
Let us denote
\begin{equation}
S(x)=\frac{i^N \Gamma^N(-i(x+i/2))}{q(x)q(x+i)},
\label{r}
\end{equation}
then
\begin{equation}
\frac{t(x)\Gamma^N(-i(x-i/2))}{q(x+i)q(x-i)}=S(x)-S(x-i).
\end{equation}
 $S(x)$ can be rewritten as follows: 
\begin{equation}
S(x)=i^N \Gamma^N(-i(x+i/2))
\biggl[\frac{q_1(x)}{q(x+i)}+\frac{q_2(x)}{q(x)}\biggr],
\end{equation}
where $q_1(x)$ and $q_2(x)$ are polynomials of degree less than $n$. 
Substituting this expansion into Baxter equation (\ref{B1}) we get:
\begin{equation}
\frac{t(x)}{q(x+i) q(x-i)}=(x-i/2)^N\biggl[
\frac{q_1(x)}{q(x+i)}+\frac{q_2(x)}{q(x)}\biggr]
+i^N\biggl[\frac{q_1(x-i)}{q(x)}+\frac{q_2(x-i)}{q(x-i)}\biggr]
\end{equation}
Since $t(x)$ is a polynomial we see that $(x-i/2)^N q_2(x)+i^N
q_1(x-i)=r(x)
q(x)$, where $r(x)$ is a polynomial with the degree less than $N$.
Expressing
then
$q_1(x)$ via $q_2(x)$ and $r(x)$, let us substitute it in the expression
for
$S(x)$:
\begin{equation}
S(x)=i^N \Gamma^N(-i(x+i/2))r(x+i)+i^N \Gamma^N(-i(x+i/2))
\frac{q_2(x)}{q(x)}
-\Gamma^N(-i(x+3 i/2)) \frac{q_2(x+i)}{q(x+i)}
\end{equation}
Now our task is to present $S(x)$ in the following form
\begin{equation}
S(x)=\frac{p(x+i)}{q(x+i)}-\frac{p(x)}{q(x)}, 
\end{equation}
and $p(x)$ will be the eigenvalue of the second $Q$-operator.
Indeed from (\ref{r}) we get 
\begin{equation}
i^N \Gamma^N(-i(x+i/2))=p(x+i) q(x)-p(x) q(x+i),
\label{Wr1}
\end{equation}
and
$i^N \Gamma^N(-i(x+i/2))^N \Gamma^N(-i(x-i/2))=p(x) q(x-i)-p(x-i) q(x)$.
Multiplying the last equation  by $(-i(x-i/2))^N$ and subtracting it from
the
previous one we see, that $p(x)$ satysfies the same Baxter equation:
\begin{equation}
t(x)p(x)=(x-i/2)^N p(x-i)+i^N p(x+i).
\end{equation}
Thus the next step is to find the function $g(x)$ such that 
\begin{equation}
g(x+i)-g(x)=i^N \Gamma^N(-i(x+i/2)) r(x+i).
\end{equation}
Let us look for $g(x)$ in the following form: 
\begin{equation}
g(x)=\sum_{k=0}^{\infty} f(-ix-k).
\end{equation}
In this case $g(x+i)-g(x)=-f(-i x)$, and we see that if $$f(-ix)=-i^N
\Gamma^N
(-i(x+i/2))
r(x+i),$$ then 
\begin{equation}
g(x)=-i^N \sum_{k=0}^{\infty} \Gamma^N(-i(x+i/2)-k) r(x+i-ik),
\end{equation}
and the desired eigenvalue will be given 
\begin{equation}
p(x)=g(x)q(x)-i^N \Gamma^N (-i(x+i/2))q_2(x)
\label{second}
\end{equation}
Apparently (\ref{second}) is a meromorphic function with respect to the
spectral parameter $x$ which
has the poles
at the integer values of $y=-ix+1/2$ (the convergency of the series for
$g(x)$
at $-ix+1/2 \neq \mathbb Z$ is provided by the term $-k$ in the
gamma function argument).

As an illustration consider a simple example for the concrete polynomial
solution of Baxter equation for the case of two degrees of freedom
$$q(x)=x^2-2ix+1/4.$$ This solution corresponds to the Bethe vector
$\frac{1}{\sqrt2}(|2,0>-|0,2>)$ and the eigenvalue of
$$t(x)=x^2-3ix-9/4.$$Here
we have introduced following notation for the vectors of quantum space: 
$$
|k_1,k_2>=(\varphi^+_1)^{k_1}(\varphi^+_2)^{k_2}|0,0>,
$$
where $|0,0>$ is the Bethe vacuum ($\varphi_1|0,0>=\varphi_2|0,0>=0$ ).\\
The explicit construction of the polynomials $q_1,~q_2,~r$ using the
method
described above gives:
$$q_1(x)=-i/2 x+1/4,~~ q_2(x)=i/2 x+3/4,~~r(x)=i/2x+1/4.$$
And for the eigenvalue of the second $Q$-operator we obtain
\begin{eqnarray}
&&p(x)=(x^2-2ix+1/4)\sum_{k=0}^{\infty} \Gamma^2(-ix+1/2-k)(i/2x
+k/2-1/4)+\nonumber\\
&&+ \Gamma^2 (-ix+1/2)(i/2~x+3/4) \nonumber
\end{eqnarray}

\section{Basic $Q$-operators for the DST model}

In the present paper we shall construct the basic $Q$-operators using the
method described in {\cite{TodaPronko}}. In this approach the
$Q^{(1,2)}$-operators are the traces of monodromies 
$\hat Q^{(1,2)}$ of appropriate $M^{(1,2)}_n$-operators acting in $n$-th
quantum
space and in the auxilliary space $\Gamma$, which is the representation
space
of Heisenberg algebra $[\rho,\rho^+]=1$. As we shall need to consider the
product of $L(x)M^{(1,2)}(x)$, the mutual auxilliary space for this object
are
$\Gamma \otimes \mathbb C^2$, where we can introduce the projectors:
\begin{equation}
\Pi_{ij}^{+}=\left(\begin{array}{c} 1\\ 
\rho \end{array}\right)_i
\frac{1}{(\rho^+\rho+1)}(1,\rho^+)_j,~~\nonumber\\
\Pi_{ij}^{-}=\left(\begin{array}{c} -\rho^+\\ 
1 \end{array}\right)_i \frac{1}{(\rho^+\rho+2)}(-\rho,1)_j
\end{equation}

According to the method of \cite{TodaPronko} we impose the condition that
the
products $ L(x) M(x)$ and $M(x) L(x)$ have  triangle forms in the sense of
projectors $\Pi^{\pm}$:
\begin{equation}
\left\{ \begin{array}{c}
\Pi_{ik}^-\left(L_{n}(x)\right)_{kl}M_{n}^{(1)}(x)\Pi_{lj}^+=0 \\
\Pi_{ik}^+M_{n}^{(1)}(x)\left(L_{n}(x)\right)_{kl}\Pi_{lj}^-=0
\end{array}, \right.
\label{tr1}
\end{equation}
for $M^{(1)}_n(x)$ and
\begin{equation}
\left\{ \begin{array}{c}
\Pi_{ik}^+\left(L_{n}(x)\right)_{kl}M_{n}^{(2)}(x)\Pi_{lj}^-=0 \\
\Pi_{ik}^-M_{n}^{(2)}(x)\left(L_{n}(x)\right)_{kl}\Pi_{lj}^+=0,
\end{array}
\right.
\label{tr2}
\end{equation}
for $M_{n}^{(2)}(x)$. Consider first the system for $M^{(1)}$. 
It follows from the first equation in (\ref{tr1}) that:
\begin{equation}
\left\{ \begin{array}{c} 
M^{(1)}(x) \left( \begin{array}{c} 1 \\ \rho 
\end{array} \right)_i =\tilde L(x)_{ij} \left( \begin{array}{c} 1 \\ \rho 
\end{array} \right)_j A^{(1)}(x) \\
B^{(1)}(x) \left( \begin{array}{c} 1 \\ \rho 
\end{array} \right)_i =\tilde L(x+i)_{ij} \left( \begin{array}{c} 1 \\
\rho 
\end{array} \right)_j M^{(1)}(x)
\end{array} \right.  ,
\label{system1}
\end{equation}
where we introduce
$$
\tilde L(x)=\left( \begin{array}{cc} i & -\varphi^+ \\
-\varphi & x-3i/2-\varphi^+ \varphi \end{array} \right),
$$
with properties $L(x)\tilde L(x)=i(x-i/2) \cdot I$
($I$ is the identity matrix) and $ L(x)+\tilde L(x+i)=Tr L(x)\cdot I$
(this
identity provides the argument shift in (\ref{system1}) and leads to the
finite
differences equation). System (\ref{system1}) has the solution of the
form
$B^{(1)}(x)=c M^{(1)}(x+i),~A^{(1)}(x)=c^{-1}M^{(1)}(x)$, where $c$ is a
number.
Let us choose $c=i$. Along with the analogous consideration of the
triangularity condition for right multiplication
$\Pi_{ik}^+M_{n}^{(1)}(x)\left(L_{n}(x)\right)_{kl}\Pi_{lj}^-=0$ it leads
to
the system:
\begin{equation}
\left\{ \begin{array}{c} 
\tilde L(x+i)_{ij} \left( \begin{array}{c} 1 \\ \rho 
\end{array} \right)_j M^{(1)}(x)=M^{(1)}(x+i) \left( \begin{array}{c} 1 \\
\rho 
\end{array} \right)_i \\
M^{(1)}(x) L(x)_{ij} \left( 
\begin{array}{c} -\rho^+ \\ 1 \end{array} \right)_j
=i\left( \begin{array}{c} -\rho^+ \\ 1 \end{array} \right)_i 
M^{(1)}(x+i)
\end{array} \right.  ,
\label{system2}
\end{equation}

For $M^{(2)}$ we get:
\begin{equation}
\left\{ \begin{array}{c} 
\tilde L(x+i)_{ij} \left( \begin{array}{c} -\rho^+ \\ 1 
\end{array} \right)_j M^{(2)}(x)=M^{(2)}(x+i) \left( \begin{array}{c}
-\rho^+
\\ 1
\end{array} \right)_i \\
M^{(2)}(x) L(x)_{ij} \left( 
\begin{array}{c} 1 \\ \rho \end{array} \right)_j
=i\left( \begin{array}{c} 1 \\ \rho \end{array} \right)_i 
M^{(2)}(x+i)
\end{array} \right.  .
\label{system3}
\end{equation}

 The full multiplication rules have the following form:
\begin{eqnarray}
&&\left(L_{n}(x)\right)_{ij}M_{n}^{(1)}(x)=\left(\begin{array}{c} 1\\ 
\rho\end{array}\right)_{i}M_{n}^{(1)}(x-i)\frac{1}{\rho^+\rho+1}(1,\rho^+)
_{j}+\nonumber\\
&+&\left(\begin{array}{c} -\rho^+\\ 1 \end{array}\right)_{i}
\frac{1}{\rho^+\rho+2}M_{n}^{(1)}(x+i)(-\rho,1)_{j}+\Pi_{ik}^+
\left(L_{n}(x)\right)_{kl}M_{n}^{(1)}(x)\Pi_{lj}^-\\
&&\left(L_{n}(x)\right)_{ij}M_{n}^{(2)}(x)=\left(\begin{array}{c} 1\\ 
\rho \end{array}\right)_{i}
\frac{1}{\rho^+\rho+1}M_{n}^{(2)}(x+i)(1,\rho^{+})_{j}+\nonumber\\
&+&\left(\begin{array}{c} -\rho^+\\ 
1\end{array}\right)_{i}M_{n}^{(2)}(x-i)
\frac{1}{\rho^+\rho+2}(-\rho,1)_{j}+\Pi_{ik}^-\left(L_{n}(x)\right)_{kl}M_
{n}^{(2)}(x)\Pi_{lj}^+ \\
&&M_{n}^{(1)}(x)\left(L_{n}(x)\right)_{ij}=\left(\begin{array}{c} 1\\ 
\rho\end{array}\right)_{i}\frac{1}{\rho^+\rho+1}M_{n}^{(1)}(x-i)(1,\rho^+)
_{j}+\nonumber\\
&+&\left(\begin{array}{c} -\rho^+\\ 1 \end{array}\right)_{i}
M_{n}^{(1)}(x+i)\frac{1}{\rho^+\rho+2}(-\rho,1)_{j}+\Pi_{ik}^-
\left(L_{n}(x)\right)_{kl}M_{n}^{(1)}(x)\Pi_{lj}^+\\
&&M_{n}^{(2)}(x)\left(L_{n}(x)\right)_{ij}=\left(\begin{array}{c} 1\\ 
\rho \end{array}\right)_{i}
M_{n}^{(2)}(x+i)\frac{1}{\rho^+\rho+1}(1,\rho^{+})_{j}+\nonumber\\
&+&\left(\begin{array}{c} -\rho^+\\ 
1\end{array}\right)_{i}\frac{1}{\rho^+\rho+2}M_{n}^{(2)}(x-i)
(-\rho,1)_{j}+\Pi_{ik}^+\left(L_{n}(x)\right)_{kl}M_
{n}^{(2)}(x)\Pi_{lj}^-\\
\nonumber
\label{right}
\end{eqnarray}

We do not consider the irrelevant structures of the last terms in the rhs
of
these rules. 
Apparently, the triangle structure (\ref{tr1},\ref{tr2}) will be valid
also
for products of $L_n$ and $M_n$, as the quantum operators with different
$n$
commute with each other, therefore these relations guarantee that both the
traces of monodromies (if they exist)
\begin{equation}
Q^{(1,2)}(x)=Tr\hat Q^{(1,2)}(x)=Tr\prod_{k=1}^{N} M^{(1,2)}_k(x),
\end{equation}
satisfy Baxter equation (\ref{Baxter}).

To solve the equations (\ref{system2},\ref{system3}) we shall use the
holomorphic representation for the operators $\rho,~ \rho^+$. Let the
operator
$\rho^+$ be the operator of multiplication:
$(\rho^+ \psi)(\alpha)=\alpha\psi(\alpha)$, while $\rho$ is the operator
of
differentiation $(\rho \psi)(\alpha)=\frac{\partial}{\partial \alpha}
\psi(\alpha)$. The action of an operator in the holomorphic representation
is
defined by its kernel:
\begin{equation}
(\hat M\psi)(\alpha)=\int d^2\mu(\beta) M(\alpha, \bar \beta) \psi(\beta),
\end{equation}
where the measure is defined as follows:
$
d^2\mu(\beta)=e^{-\beta \bar \beta} d \beta d \bar \beta.
$

In this representation  the operators, which satisfy the systems
(\ref{system2},\ref{system3}) have the following forms:
\begin{eqnarray}
M^{(1)}(x,\alpha,\bar \beta)=e^{-i \bar\beta \varphi^+}
\frac{\Gamma(-i(x-i/2))}
{\Gamma(-\varphi^+\varphi-i(x-i/2))} e^{-i \alpha \varphi} \nonumber\\
M^{(2)}(x,\alpha,\bar\beta)=e^{i \alpha \varphi}
e^{i \pi \varphi^+\varphi} \Gamma(-\varphi^+ \varphi-i(x-i/2))
e^{i \bar\beta \varphi^+}
\label{Mkern}
\end{eqnarray}

In order to find the monodromy $\hat Q(x,\alpha,\bar \beta)^{(1,2)}$ one
has to
perform an ordered multiplication of $M^{(1,2)}$- operators:

\begin{eqnarray} 
\hat Q^{(i)}(x,\alpha,\bar\beta)=\int
\prod_{i=1}^{N-1}d^2\mu(\gamma_{i})
M^{(i)}_{N}(x,\alpha,\bar\gamma_{N-1} )
M^{(i)}_{N-1}(x,\gamma_{N-1},\bar\gamma_{N-2})\\
\cdots\times M^{(i)}_{2}(x,\gamma_{2},\bar\gamma_{1})
M^{(i)}_{1}(x,\gamma_{1},\bar\beta).
\end{eqnarray}

Taking the trace of $\hat Q^{(1,2)}$ over the auxiliary space we obtain
$Q^{(1,2)}-$operators. The trace of an operator $Q$
in the holomorphic representation is given by 
\begin{equation}
TrQ=\int{\mbox{d}^2\mu(\alpha)\hat Q(\alpha,\bar \alpha)},
\end{equation}
where $\hat Q(\alpha,\hat \alpha)$ is the kernel of $\hat Q$.

The eigevalues $Q^{(1)}(x)$ are polinomials in the spectral parameter $x$.
It can be seen from the action of $Q^{(1)}$ onto the basic vectors
$|~n_1,n_2,...,
n_N>=(\varphi^{+}_1)^{n_1}(\varphi^{+}_2)^{n_2}...(\varphi^{+}_N)
^{n_N}|0>,$ where $|0>$ is the Bethe vacuum: $\varphi_{k}|0>=0,~k=1..N$: 
\begin{eqnarray}
&&Q^{(1)}(x)|n_1,..., n_N>=\sum\limits_{m_1,...,m_N=0}^{n_1,...,n_N}
\prod_{k=1}^{N}\frac{(-1)^{m_{k}}}{m_{k}!}\nonumber\\
&&\frac{\Gamma(-i(x-i/2))}{\Gamma(-i(x-i/2)-n_k+m_k)}
\frac{n_k!}{(n_k-m_k)!}|..., n_k-m_k+m_{k-1},...>
\label{Q1acts}
\end{eqnarray}
We see $Q^{(1)}(x)$ leaves the subspace of vectors with a common particle
number
$n=n_1+n_2+...+n_N$ invariant, and all matrix elements of $Q^{(1)}$ are
polinomials in $x$. We shall see below that
$[Q^{(1)}(x_1),Q^{(1)}(x_2)]=0$, so the eigenvalues of $Q^{(1)}$ are
polinomials in $x$ too.
Constructed in \cite{SklyaninDST} $Q$-operator corresponds to $Q^{(1)}$.
Its action onto the basic vectors (in the paper \cite{SklyaninDST} the
coordinate representation has been choosen for the quantum operators with
the
basic vectors: $x_1^{n_1} \dots x_n^{n_N}$) is similar to $Q^{(1)}$ in
(\ref{Q1acts}).

For comparison we give also the action of $Q^{(2)}$ onto the same basic
vectors:
\begin{eqnarray}
&&Q^{(2)}(x)|n_1,...,n_N>=\nonumber\\
&&=e^{i \pi n} \cdot \sum \limits_{m_1,...,m_N=0}^{\infty} \prod
\limits_{k=1}^{N} \Gamma(-ix-1/2-n_k-m_{k-1}) \cdot \nonumber\\
&&\cdot \frac{(m_{k-1}+n_k)!}{m_k!(n_k+m_{k-1}-m_k)!}
|...,n_k+m_{k-1}-m_k,...>.
\label{Q2acts}
\end{eqnarray}
Here the summation in contrast to $Q^{(1)}$ is taken over an infinite set
of
$m_k$,  restricted however by the condintions $m_k-m_{k-1}\leq n_k$. 

Let us notice that in some realizations of quantum and axilliary operators
there may appears the factorization of $Q$-operators first considered by
Bazhanov and Stroganov \cite{BazhStr} (see also \cite{PasqGod}), if, for
example, we choose the
coordinate representaion for quantum and auxiliary operators, we will get
the
following factorized form for $Q$-operator \cite{SklyaninDST,PasqGod}:
\begin{equation}
Q(x_1,...,x_N,{x'}_1,...,{x'}_N)=\prod\limits_{k=1}^{k=N}q_k(x_k,x'_{k+
1},{x'}_k)
\label{factor}
\end{equation}
In the paper \cite{SklyaninDST} one of the $Q$-operators in
the form (\ref{factor}) was constructed. It is also 
possible to construct the second $Q$-operator (it was also notice
in \cite{SklyaninDST}) in the same factorized form without use of an
axilliary
space. However from the point of view of approach \cite{TodaPronko} the
origin
of such kind of factorization is not clear.
and
trace of
factorized
defined by
in
complex
canonical
onto
also
above
$[Q(x),t(y)]=0$,
these

In the simplest case of one quantum degree of freedom $N=1$ we obtain
($n=\varphi^+\varphi$)
\begin{eqnarray}
Q^{(1)}(x)=\sum_{k=0}^{n}\frac{n!}{k!(n-k)!}\frac{\Gamma(-ix+1/2)}{\Gamma(
-ix+1
/2-n-k)}\\
Q^{(2)}(x)=e^{i\pi n}\sum_{m=0}^{\infty}\frac{(n+m)!}{m!n!}
\Gamma(-ix-1/2-n-m).
\end{eqnarray}

As we have expected the eigenvalues of $Q^{(1)}$ are polynomial of degree
$n$
and the eigenvalues of $Q^{(2)}$ are meromorphic functions over the
spectral
parameter.

 It is possible to find the solutions of
(\ref{system2},\ref{system3}) in the form:
\begin{eqnarray}
&&M^{(1)}(x,\rho)=\mbox{P}^{\rho\varphi}(i-\varphi
\rho^+)^{-i(x-i/2)}e^{-x
\pi/2}\nonumber\\
&&M^{(2)}(x,\rho)=\Pi^{\rho\varphi}\Gamma(-\rho^+\rho-\varphi^+\varphi-i(x
-i/2)
)
\end{eqnarray}
Where $\mbox{P}^{\rho\varphi}$ is an operator equal:
$$
\mbox{P}^{\rho\varphi}=exp\left[\pi/2(\varphi^+ \rho-\varphi \rho^+
)\right]
exp\left[i\pi/2(\rho^+ \rho-\varphi^+\varphi)\right].
$$
It has properties of the permutation operator:
\begin{eqnarray}
&&\mbox{P}^{\rho\varphi}\varphi=i\rho\mbox{P}^{\rho\varphi},~~
\mbox{P}^{\rho\varphi}\varphi^+=-i\rho^+\mbox{P}^{\rho\varphi},
\nonumber\\
&&\varphi\mbox{P}^{\rho\varphi}=-i\mbox{P}^{\rho\varphi}\rho,~~
\varphi\mbox{P}^{\rho\varphi^+}=i\mbox{P}^{\rho\varphi}\rho^+,
\end{eqnarray}
 $\Pi^{\rho\varphi}$ is an operator with the following properties:
\begin{eqnarray}
&&\varphi\Pi^{\rho\varphi}=-i\rho\Pi^{\rho\varphi},~~
\Pi^{\rho\varphi}\varphi^+=-i\Pi^{\rho\varphi}\rho^+, \nonumber\\
\mbox{however} \nonumber\\
&&\varphi^+\Pi^{\rho,\varphi}=i[\Pi^{\rho,\varphi},\rho^+],~~
\Pi^{\rho,\varphi}\varphi=-i[\Pi^{\rho,\varphi},\rho]. \nonumber 
\end{eqnarray}
 The explicit expression for $\Pi^{\rho,\varphi}$ is
\begin{eqnarray}
\Pi^{\rho\varphi}=&&\left[1+\sum_{k=1}(i\varphi\rho^+)^k\frac{\Gamma(\rho^
+\rho
+1)}{\Gamma(\rho^+\rho+k+1)}+
\sum_{k=1}(i\varphi^+\rho)^k
\frac{\Gamma(\varphi^+\varphi+1)}{\Gamma(\varphi^+\varphi+k+1)}
\right]\cdot\nonumber\\
&&\cdot\frac{\Gamma(\rho^+\rho+\varphi^+\varphi+1)}{\Gamma(\rho^+\rho+1)
\Gamma(
\varphi^+\varphi+1)}e^{i\pi\varphi^+\varphi}.
\end{eqnarray}

The operators $M_n^{(1,2)}$ and $L_n(x)$ satisfy certain intertwining
relations
which will imply the mutual commutativity of $Q$-operators and the
transfer
matrix. Here we will present the intertwining relations without its
derivation
because the method used and the intertwining $R$-matrix are the same as in
the
case of the Toda chain
{\cite{TodaPronko}}.

They are:
1) for the Lax and $M$-operators:
\begin{equation}
R^{(i)}_{kl}(x-y)\left(L_{n}(x)\right)_{lm}M_{n}^{(i)}(y)=M_{n}^{(i)}(y)
\left(L_{n}(x)\right)_{kl}R^{(i)}_{lm}(x-y)
\label{MT}
\end{equation}
The corresponding $R$-matrices are:
\begin{eqnarray}
R^{(1)}_{kl}(x-y)=\left(\begin{array}{cc} x-y+i\rho^+\rho&-i\rho^+\\
-i\rho&i  \end{array}\right)\\
\nonumber\\
R^{(2)}_{kl}(x-y)=\left(\begin{array}{cc} i&i\rho^+\\
i\rho&x-y+i+i\rho^+\rho  \end{array}\right)
\end{eqnarray} 
Note that in the paper \cite{SklyaninDST} the equation (\ref{MT}) with
$R^{(1)}$ was considered as the defining equation for local $M$-operators
and
the trace
of their monodromy is  the $Q$-operator.

2) for  $M^{(1)}$- and $M^{(2)}$-operators acting in different auxiliary
spaces
\begin{equation}
M_{n}^{(1)}(x,\rho)M_{n}^{(2)}(y,\tau)R^{12}(x-y)=R^{12}(x-y)M_{n}^{(2)}(y
,\tau)M_{n}^{(1)}(x,\rho),
\end{equation}
where the intertwining matrix $R^{12}$ is defined by the kernel in the
holomorphic representation
\begin{equation}
R^{12}(x,\alpha,\bar\beta;\gamma,\bar\delta)=\sum_{n=0}\frac{\left((\alpha
-\gamma)(\bar\beta-\bar\delta)\right)^n}{n!\Gamma(-ix+n+1)},
\end{equation} 
here $\alpha,\bar\beta$ and $\gamma, \bar\delta$ are the holomorphic
variables
for the representation space for pairs of operators $\rho, \rho^+$ and
$\tau,
\tau^+$.

3) for $M^{(1)}$-operators acting in different auxiliary spaces:  
\begin{equation}
R^{(11)}(x-y)M^{(1)}(x,\rho)M^{(1)}(y,\tau)=M^{(1)}(y,\tau)M^{(1)}(x,\rho)
R^{(11)}(x-y),
\end{equation}
where
\begin{equation}
R^{(11)}(x)=P_{\rho\tau}(1+\rho^+\tau)^{-ix},
\end{equation}
here $P_{\rho\tau}$ is the operator of permutation of the auxiliary
spaces.

4) for $M^{(2)}$-operators acting in different auxiliary spaces:
\begin{equation}
R^{(22)}(x-y)M^{(2)}(x,\rho)M^{(2)}(y,\tau)=M^{(2)}(y,\tau)M^{(2)}(x,\rho)
R^{(22)}(x-y),
\end{equation}
where
\begin{equation}
R^{(22)}(x)=P_{\rho\tau}(1+\tau^+\rho)^{-ix}.
\end{equation}

These intertwining relations lead to the mutual commutativity of
$Q$-operators
and $t$-matrix.
\begin{equation}
[t(x),Q^{(i)}(y)]=0, \quad [Q^{(i)}(x),Q^{(j)}(y)]=0, \quad i,j=1,2.
\end{equation}

So far we have constructed two solutions of the
operator Baxter equation. Now we are going  to establish linear
independence of
these solutions defining the  one-parametric family of a finite
difference analogues of the Wronsky determinant
\begin{equation}
W_m=Q_{1}(x-im)Q_{2}(x+i)-Q_{1}(x+i)Q_{2}(x-im)
\end{equation}
where $m$ is a non-negative integer. Consider properties of these objects
following directly from Baxter equation:
\begin{eqnarray}
&&t(x)Q_1(x)=(x-i/2)^N Q_1(x-i)+i^N Q_1(x+i) \nonumber\\
&&t(x)Q_2(x)=(x-i/2)^N Q_2(x-i)+i^N Q_2(x+i). 
\end{eqnarray}
Multiplying first equation by $Q_2(x)$, second by $Q_1(x)$, and
substracting
one from another we see that
$$
(x-i/2)^N W_0(x-i)=-i^NW_0(x),
$$
so $W_0$ necessarily has the factor $\Gamma^N(-i(x+i/2))$. Multiplying
then
the first equation by $Q_2(x-i)$ and the second  by $Q_1(x-i)$ we get
$$
W_1(x)=(-i)^N t(x) W_0(x-i).
$$
In the general case of non-negative integer $m$ multiplying the first
equation
by  $Q_2(x-im)$ and the second equation by $Q_1(x-im)$, we obtain:
\begin{equation}
(x-i/2)^N W_{m-2}(x-2i)+i^N W_m(x)=t(x)W_{m-1}(x-i).
\label{fusion}
\end{equation}
 These identities completely define all $W_m$ provided $W_0$ is known.
These argumentations make sense in the presence of two solutions of Baxter
equation
with $W_0\neq0$ identically.
The eigenvalues of the transfer matricies $t_{l}(x)$-traces of monodromy
matrixes in auxiliary spaces of spin $l=m/2$ satisfies the recurrent
relations
similar to (\ref{fusion}). The family of $t_l$ could be obtained with the
help
of the expression for the Lax operator in the auxiliary space of spin $l$:
\begin{equation}
L_l(x)=i^{2l}e^{-il^+\varphi^+}\frac{\Gamma(l^3-ix)}
{\Gamma(-ix-l)}e^{-il^-\varphi}.
\end{equation}
Here operators $l^{k}~(k=\pm,~3)$ are the operators of spin $l$ and the
factor
$i^{2l}\Gamma^{-1}(-i(x+i/2))$
is introduced in order that in the cases of $l=0$ and $l=1/2$ we will
obtain
correspondingly $L_0=1$ and $L_{1/2}(x)=L(x)$-the Lax operator
(\ref{Lax}).
For the operators considered in the Introduction the relation (\ref{Wr1})
gives
$$W_0(x)=i^N\Gamma^N(-i(x+i/2)).$$ An explicit calculation of $W_0$ for
the
solutions constructed in Section 2 using the method, described in the
paper
\cite{TodaPronko} gives
\begin{equation}
W_0(x)=e^{i\pi \hat n}\Gamma^N(-i(x+i/2)),
\end{equation}
where $\hat n$ is the number of particles operator. And for this pair it
follows that:
\begin{equation}
W_m(x)=e^{i\pi \hat n} (-i)^{2lN}\Gamma^N(-ix-1/2) t_l(x)
\end{equation}
Finally we arrive at the following general Wronskian-type relations:
\begin{equation}
Q_{1}(x-im)Q_{2}(x+i)-Q_{1}(x+i)Q_{2}(x-im)=e^{i\pi \hat n}
(-i)^{2lN}\Gamma^N(-ix-1/2) t_l(x).
\end{equation}
\section{Conclusion}

In the present paper we have constructed the basic $Q$-operators in the
form
of traces of monodromies of basic local $M$-operators for the case of DST
integrable
model. The obtained $Q$-operators are presented in the form of formal
series
over
the canonical operators $\varphi_k, \varphi_k^+$ and have well defined
action
onto vectors of quantum space. The intertwining relations indicating the
mutual
commutativity of $t(x)$ and $Q$-operators are derived. Obtained are
functional
relations of wronskian-type showing the linear independence of
$Q$-operators
and connection between the $Q$-operators and the transfer matrices in the
auxiliary spaces of higher spins. 

Let us notice some unsolved problems. It would be interesting to find
$Q$-operators for small numbers of freedom degrees as functions of the
family
of commuting operators, connected with $t(x)$. The origin of the
factorizations 
a l\'a Pasquer-Gaudin, which appear in some representations
\cite{PasqGod, SklyaninDST} is
not
clear. 

The described in the present method makes it possible to find
$M$-operators in
the most interesting case of the XXX-spin chain (they coincide with the
Lax
operators $L(x)$ and $\tilde L(x)$ for DST model with interchaged quantum
and
auxiliary spaces), but the traces of their monodromy diverge. However,
there
exists the procedure of $Q$-operator construction, analogous to the
described
in \cite{PasqGod} one, for XXX SL(2,$\mathbb C$) spin chain
\cite{DerkXXX}. 
So the case of XXX-spin chain deserves further investigation.

This work was supported in part by grants of RFBR 00-15-96645,
01-02-16585, 
CRDF MO-011-0, of the Russian Minestry on the education E00-3.3-62 and
INTAS
00-00561.

\end{document}